# Graphitic-$C_3N_4$/$TiO_2$(B) S-scheme Heterojunctions for Efficient Photocatalytic $H_2$ Production and Organic Pollution Degradation

Xiaoyi Zhou[1], Min Zhang[1,*], Qiushi Wang[2], Shiwen Du[2], Xuedong Jing[2], Zhenyi Zhang[2,*]

1. School of Physics & Electronic Technology, Liaoning Normal University, Dalian, 116029, China

2. School of Physics and Materials Engineering, Dalian Minzu University, Dalian 116600, China

**Abstract**

Achieving both broad solar-spectrum absorption and strong redox capability is critical for semiconductor photocatalysts in environmental remediation and energy conversion. Herein, an S-scheme heterojunction photocatalyst is constructed by coupling $TiO_2$(B) nanorods with g-$C_3N_4$ nanosheets. Its well-matched band structure extends light absorption from the UV to the visible region and enables efficient charge separation. Under simulated sunlight irradiation, the 40 wt% g-$C_3N_4$/$TiO_2$(B) heterojunction delivers a $H_2$ evolution rate of 1.98 mmol g$^{-1}$ h$^{-1}$ for water reduction with methanol as the sacrificial agent, which is 1.5 and 2.0 times higher than those of pure g-$C_3N_4$ and $TiO_2$(B), respectively. When exposed to amoxicillin wastewater instead of methanol solution, the heterojunction degrades 98.2% of amoxicillin and produces 20.70 μmol g$^{-1}$ of $H_2$ within 90 min. Moreover, the heterojunction shows excellent photodegradation activity toward various organic antibiotics and dyes, owing to the S-scheme charge separation mechanism. This work highlights the promising potential of S-scheme heterojunctions for photocatalytic $H_2$ production coupled with organic wastewater treatment.

* Corresponding authors, Email: m.zhang@live.com; zhengzy@dlnu.edu.cn;

**Keywords:** Photocatalysis; S-Scheme Heterojunction; TiO$_2$/g-C$_3$N$_4$ Composite; Hydrogen Evolution; Pollutant Degradation

# 1. Introduction

With the rapid advancement of global industrialization and the healthcare sector, the releasing of organic wastewater has escalated into a pressing environmental concern due to its detrimental effects on aquatic ecosystems and human health [1–4]. Conventional treatment techniques, such as adsorption and biological methods, are often constrained by significant drawbacks, including high operational costs, limited resource recovery, and the risk of secondary pollution [5,6]. Consequently, the development of efficient, sustainable, and resource-oriented strategies for organic wastewater treatment is of paramount importance. Among various emerging technologies, semiconductor-based photocatalysis has garnered considerable attention, owing to its ability to harness solar energy to drive redox reactions for both the oxidative degradation of organic contaminants and the photoreduction of water to produce hydrogen gas (H$_2$) [7]. Despite these advantages, the practical deployment of single-component semiconductor photocatalysts remains hindered by intrinsic limitations such as insufficient redox potential, limited absorption within the solar spectrum, and rapid recombination of photogenerated electron–hole pairs [8].

Constructing semiconductor heterojunctions by coupling two distinct semiconductor components is widely recognized as an effective strategy to overcome the intrinsic limitations of single-component photocatalysts [9]. Among the various heterojunction architectures, the S-scheme heterojunction has attracted considerable attention due to its unique charge-transfer mechanism. Typically, it consists of an oxidation photocatalyst (OP) and a reduction photocatalyst (RP). In such systems, the conduction band (CB) and valence band (VB) potentials of the RP are more negative than those of the OP, while the OP possesses a higher work function than the RP [10]. Upon contact between the two semiconductors, electrons transfer from the RP to the OP until their Fermi levels reach equilibrium, leading to the formation of an internal electric field (IEF) directed from the RP to the OP at the heterointerface. Under light irradiation, this IEF facilitates the recombination of photogenerated electrons in the CB of the OP with photogenerated holes in the VB of the RP. As a result, highly reductive electrons are preserved in the CB of the RP, while strongly oxidative holes remain in the VB of the OP [11]. Consequently, compared with single-component photocatalysts, S-scheme heterojunction photocatalysts exhibit enhanced redox capability, broadened light absorption, and more efficient separation of photogenerated charge carriers [12].

To date, numerous S-scheme heterojunction photocatalysts have been explored for photocatalytic environmental remediation and solar energy conversion [13]. Among them, systems composed of graphitic carbon nitride (g-$C_3N_4$) as the reduction photocatalyst (RP) and $TiO_2$ as the oxidation photocatalyst (OP) have attracted considerable attention because of their chemical stability, low toxicity, low cost, and well-matched band structures with complementary redox potentials [14]. Anatase and rutile $TiO_2$ are commonly employed as OPs owing to their facile synthesis and well-established photocatalytic properties. In contrast, bronze-phase $TiO_2$ ($TiO_2$(B)) exhibits superior charge-separation performance due to its anisotropic crystal structure. $TiO_2$(B) possesses a monoclinic C2/m framework composed of edge- and corner-sharing $TiO_6$ octahedra, forming open channels along the *b*-axis between axial oxygen atoms. This unique architecture facilitates directional charge transport and thereby promotes the separation of photogenerated charge carriers [15]. Therefore, constructing g-$C_3N_4$/$TiO_2$(B) S-scheme heterojunctions is of considerable interest for resource-oriented photocatalytic treatment of organic wastewater [16–19].

In this work, g-$C_3N_4$/$TiO_2$(B) S-scheme heterojunctions with different component ratios were synthesized through a three-step process (Figure 1): (i) preparation of g-$C_3N_4$ nanosheets via thermal polymerization; (ii) synthesis of $TiO_2$(B) nanorods by a hydrothermal method followed by annealing; and (iii) construction of the g-$C_3N_4$/$TiO_2$(B) heterojunction through grinding-assisted mixing and subsequent secondary annealing. Under simulated sunlight irradiation, all prepared heterojunctions exhibited enhanced photocatalytic activity for hydrogen evolution from water compared with pristine $TiO_2$(B) and g-$C_3N_4$, using methanol as a sacrificial agent. The optimal heterojunction containing 40 wt.% g-$C_3N_4$ achieved a $H_2$ evolution rate of 1.98 mmol·g$^{-1}$·h$^{-1}$, which was 2.0 and 1.5 times higher than those of pure $TiO_2$(B) and g-$C_3N_4$, respectively. When methanol was replaced by aqueous solutions containing various organic pollutants (amoxicillin, ciprofloxacin, ofloxacin, tetracycline, rhodamine B, and methyl orange), the heterojunction exhibited excellent photocatalytic degradation performance, achieving degradation efficiencies above 80% within 90 min of simulated sunlight irradiation. Notably, simultaneous $H_2$ evolution (20.70 μmol·g$^{-1}$) and amoxicillin degradation (98.2%) were achieved over the 40 wt.% g-$C_3N_4$/$TiO_2$(B) heterojunction after 90 min of irradiation. These results demonstrate that the constructed g-$C_3N_4$/$TiO_2$(B) S-scheme heterojunction enables the synergistic integration of efficient organic pollutant degradation and photocatalytic hydrogen production.

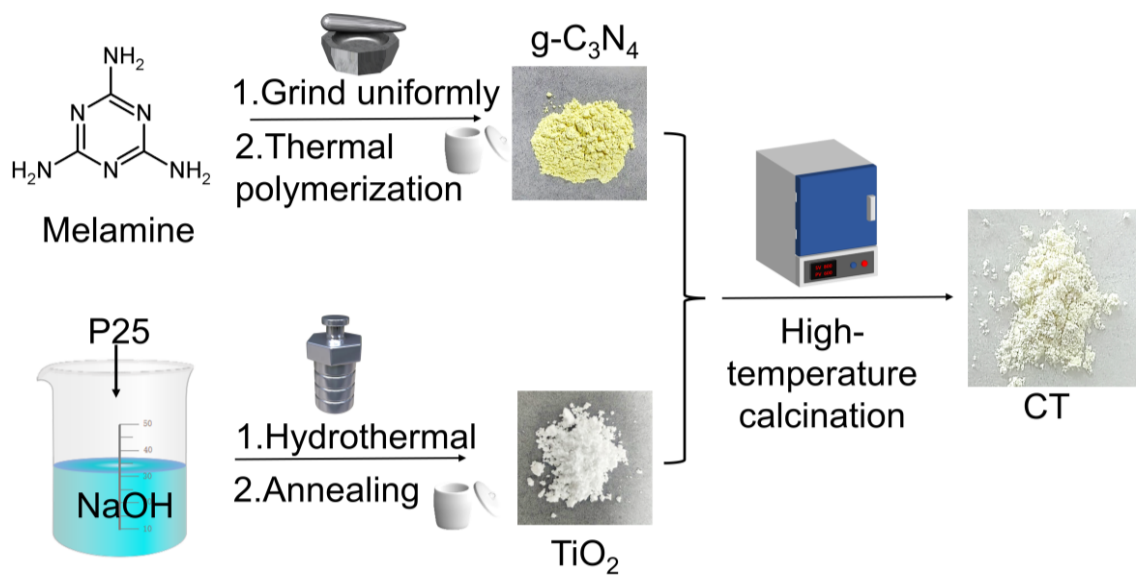

**Figure 1.** Schematic illustration of the three-step synthesis process for TiO$_2$/g-C$_3$N$_4$ heterojunction composite.

## 2. Experimental details

**Synthesis of g-C$_3$N$_4$ nanomaterials:** g-C$_3$N$_4$ is prepared through the traditional thermal polymerization process. First, the melamine powder is ground in a agate mortar for 30 minutes. Then, the fine powder (10 grams) is placed in a covered corundum crucible. Subsequently, the crucible is moved into a muffle furnace, where the temperature is raised to 500°C within 40 minutes, then further increased to 550°C within 20 minutes, and maintained for 2 hours. After natural cooling to 25°C, the light yellow g-C$_3$N$_4$ blocks are obtained. The prepared g-C$_3$N$_4$ is further ground into fine powder, which can be used as the precursor for the subsequent steps.

**Synthesis of TiO$_2$ nanomaterials:** TiO$_2$ is prepared through the traditional hydrothermal reaction. First, P25 powder (1g) is mixed with 30 ml of 10 M NaOH solution and subjected to 1 hour of ultrasonic treatment. Then, the mixture is transferred to a reaction vessel and heated at 200°C for 24 hours. After the treatment, the reaction vessel is allowed to cool naturally to room temperature. The product is collected, washed several times with 0.1 M dilute hydrochloric acid until the PH value of the solution is approximately 2, and then washed several times with deionized water until the pH value of the solution is approximately 7. The white product is centrifuged and dried at 80°C in the air for 12 hours to obtain a white block. The white block is further ground into powder, and then the fine powder is placed in a crucible and moved into the muffle furnace, where the temperature is raised to 400°C within 40 minutes and maintained for 2 hours. After natural cooling, white $TiO_2$ powder is obtained and will be used as the precursor for the subsequent steps.

**Electrochemical characterization:** The photocurrent response and electrochemical properties of the samples were investigated using an electrochemical workstation. The working electrode was prepared as follows: the photocatalyst and 0.5 mL of Nafion solution were dispersed in a water/ethanol mixture (1:1, v/v) and ultrasonicated for 30 min to obtain a homogeneous suspension. The resulting slurry was deposited onto a 1 × 1 cm² fluorine-doped tin oxide (FTO) glass substrate via dip-coating and subsequently dried at 100 °C for 12 h. A 0.5 M $Na_2SO_4$ aqueous solution was used as the electrolyte. Simulated sunlight irradiation was applied during the measurements. Photocatalytic hydrogen evolution experiments were carried out in a sealed photocatalytic reaction vessel under simulated sunlight irradiation (AM 1.5 filter) at room temperature. In a typical experiment, 10 mg of photocatalyst was dispersed in 10 mL of aqueous solution containing 2 mL of Pt precursor solution (co-catalyst), 1 mL of methanol as a sacrificial agent, and 7 mL of deionized water. Prior to irradiation, the suspension was purged with argon gas for 15 min to remove dissolved oxygen and ensure an inert reaction environment. The evolved gas was collected at regular intervals and quantified using a gas chromatograph equipped with helium as the carrier gas. The photocatalytic degradation performance of the nanocomposite was evaluated under simulated sunlight irradiation using various organic pollutants as model contaminants. In a typical test, 10 mg of photocatalyst was dispersed in 10 mL of pollutant solution with an initial concentration of 10 mg·L$^{-1}$. Before irradiation, the suspension was purged with argon for 15 min to remove dissolved oxygen. A xenon lamp equipped with an AM 1.5 filter served as the light source, positioned 10 cm from the reactor. During the reaction, 1 mL aliquots were withdrawn at regular intervals and analyzed using a Shimadzu UV-1800 UV–Vis spectrophotometer to monitor the change in pollutant concentration. The degradation efficiency was calculated. It should be noted that the photocatalytic reactions are primarily driven by visible light, as the ultraviolet fraction in the solar spectrum is relatively small. Based on the designed bandgap structure, the synthesized nanomaterials exhibit effective catalytic activity within the visible-light region.

## 3. Results and discussion

The as-prepared g-$C_3N_4$/$TiO_2$(B) heterojunction was denoted as CT. The crystal structures of the synthesized samples were characterized by X-ray diffraction (XRD). As shown in Figure 2a, the diffraction peaks of pristine $TiO_2$(B) match well with those of bronze-phase $TiO_2$ possessing a monoclinic C2/m structure [20]. For g-$C_3N_4$, two characteristic diffraction peaks located at 13.0° and 27.3° are observed, corresponding to the (100) and (002) planes of its graphitic structure, respectively [21]. These peaks are associated with the in-plane structural packing of tri-s-triazine units and the interlayer stacking of the conjugated aromatic layers. After coupling the two semiconductors, the characteristic diffraction peaks of both $TiO_2$(B) and g-$C_3N_4$ are retained in the XRD pattern of CT, indicating the successful formation of the heterojunction. Although partial overlap occurs between their (002) reflections, the major diffraction features of both components remain clearly distinguishable.

Fourier transform infrared (FT-IR) spectroscopy was employed to further investigate the chemical structure and interfacial interactions between $TiO_2(B)$ and $g-C_3N_4$ in the heterojunction (Figure 2b). For pristine $TiO_2(B)$, two absorption bands in the range of 580–860 cm$^{-1}$ are attributed to the stretching vibrations of Ti–O–Ti bonds [22]. The bands located at ~1640 and ~3400 cm$^{-1}$ correspond to the bending vibration of adsorbed water and the stretching vibration of surface hydroxyl groups, respectively [3]. For pristine $g-C_3N_4$, a broad band centered at around 3200 cm$^{-1}$ is assigned to N–H stretching vibrations. Multiple bands in the range of 1200–1600 cm$^{-1}$ originate from the characteristic stretching modes of CN heterocycles within the $g-C_3N_4$ framework [23]. In addition, the sharp peak at 808 cm$^{-1}$ is attributed to the out-of-plane bending vibration of triazine ring units [24]. All characteristic absorption bands of both $TiO_2(B)$ and $g-C_3N_4$ are preserved in the spectrum of CT, confirming the successful formation of the composite heterojunction. Notably, the triazine-ring vibration peak shifts slightly from 808 to 809 cm$^{-1}$ after coupling with $TiO_2(B)$, suggesting the presence of interfacial chemical interactions between the two components [25].

The morphology and microstructure of the synthesized samples were further examined by scanning electron microscopy (SEM) at different magnifications. As shown in Figure 2c, pristine $g-C_3N_4$ exhibits a typical layered morphology composed of aggregated nanosheets. In contrast, numerous rod-like structures with diameters below 1 μm and lengths of several tens of micrometers are observed for $TiO_2(B)$ (Figure 2d). These nanorods are randomly distributed and possess relatively smooth surfaces. In the SEM images of CT, both sheet-like $g-C_3N_4$ and rod-like $TiO_2(B)$ structures are clearly observed, further confirming the successful construction of the g-C$_3$N$_4$/TiO$_2$(B) heterojunction composite.

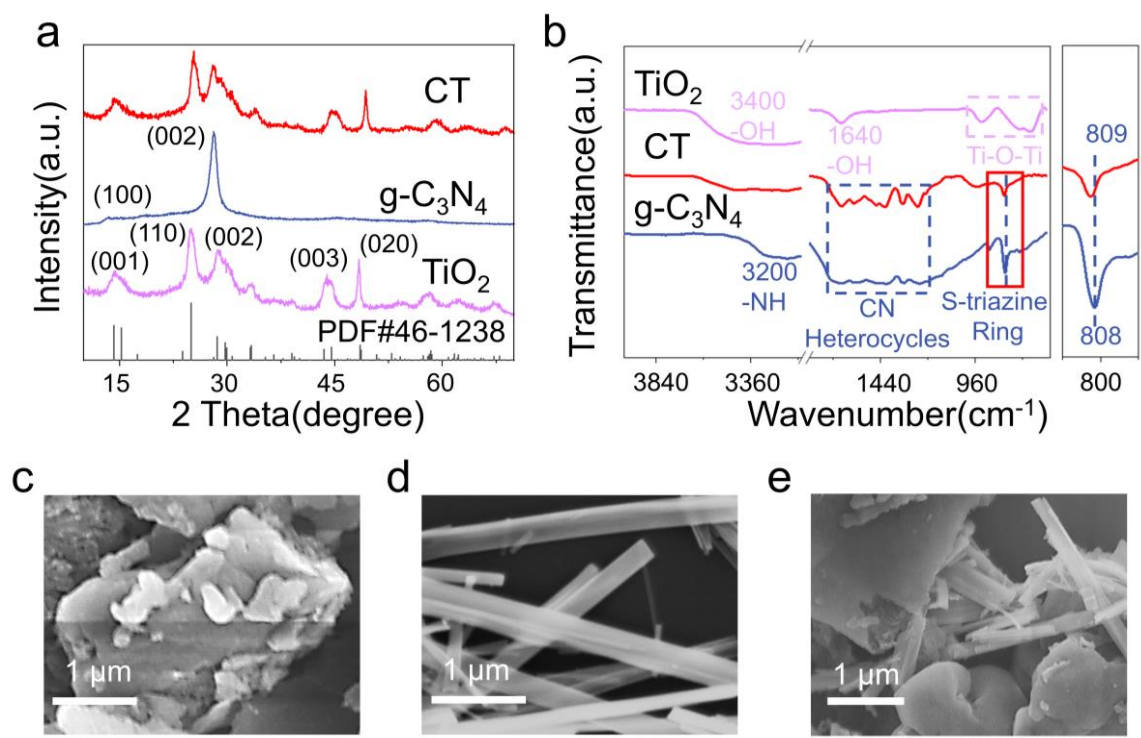

**Figure 2.** Composition and structural characterizations of the samples: (a) XRD patterns of g-C$_3$N$_4$, TiO$_2$ and CT, (b) FTIR spectra g-C$_3$N$_4$, TiO$_2$ and CT, and scanning electron microscopy (SEM) images: (c) g-C$_3$N$_4$ (d) TiO$_2$ (e) CT.

To further elucidate the microstructure of the as-prepared CT, transmission electron microscopy (TEM) was performed. The TEM images (Figure 3) reveal a composite morphology consisting of sheet- and rod-like structures, which, based on SEM observations, are tentatively assigned to g-C$_3$N$_4$ and TiO$_2$(B), respectively. High-resolution TEM (HRTEM) images provide detailed insights into the heterojunction interface. Lattice fringes with a spacing of 0.31 nm are observed, corresponding to the (002) plane of TiO$_2$(B) [26], whereas adjacent regions lacking distinct lattice fringes are attributed to the low crystallinity of g-C$_3$N$_4$. Elemental mapping (Figure 4c) further confirms the heterogeneous distribution of elements within CT. TiO$_2$(B) rods are closely adhered to g-C$_3$N$_4$ sheets, with N and C primarily localized in the sheet regions and Ti and O concentrated in the rod regions. Weak O signals detected in the g-C$_3$N$_4$ sheets are ascribed to surface oxygen impurities [27]. These observations collectively demonstrate the intimate interfacial contact between g-C₃N₄ and TiO₂(B) in the heterojunction.

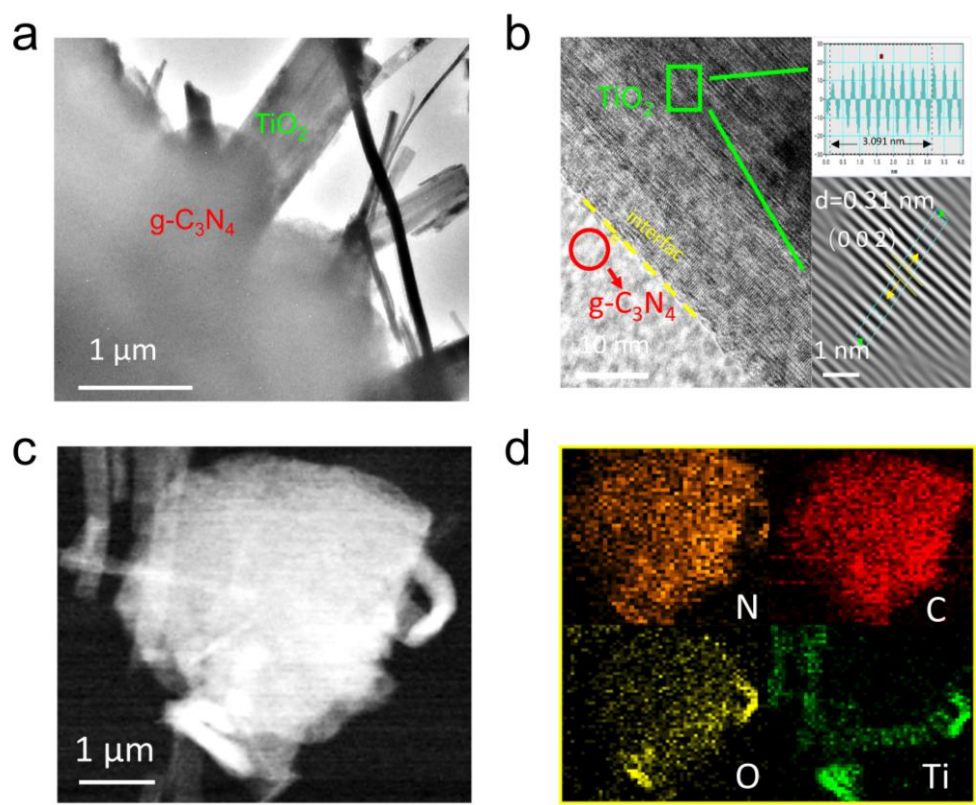

**Figure 3.** (a) HRTEM image of CT, (b) calculation of interplanar spacing through SAED, (c-d) HAADF-STEM and corresponding elemental distribution images.

X-ray photoelectron spectroscopy (XPS) was employed to investigate the surface elemental composition and chemical states of the as-prepared CT. The full survey spectrum (Figure S1, Supporting Information) confirms the presence of Ti, O, C, and N, indicating the successful incorporation of these elements into the heterojunction [28]. The high-resolution C 1s spectrum (Figure 4a) can be deconvoluted into three peaks at 284.6, 285.9, and 288.1 eV, corresponding to sp² C–C bonds, sp³ carbon bonds, and sp²-bonded carbon in N–C=N groups of g-C₃N₄, respectively [29]. The N 1s spectrum (Figure 4b) exhibits three peaks at 398.2, 400.0, and 403.7 eV, assignable to C=N–C, N–C₃, and C–N–H bonds, respectively [22]. In the O 1s spectrum, peaks at 529.5 and 531.4 eV are attributed to lattice oxygen (Ti–O–Ti) and surface hydroxyl groups (O–H) of TiO$_2$, respectively [30]. The Ti 2p spectrum displays two peaks at 458.1 and 463.8 eV, corresponding to the Ti 2p$_{3/2}$ and 2p$_{1/2}$ orbitals of Ti$^{4+}$ [31]. Notably, in the CT composite, the binding energies of C and N are slightly lower than those in pristine g-C$_3$N$_4$, whereas the Ti and O binding energies are slightly higher than those in pristine TiO$_2$(B). These shifts indicate interfacial electron transfer from g-C$_3$N$_4$ to TiO$_2$(B), consistent with the differences in their Fermi levels (E$_F$) [32]. Collectively, these XPS results further confirm the successful formation of the g-C$_3$N$_4$/TiO$_2$(B) S-scheme heterojunction.

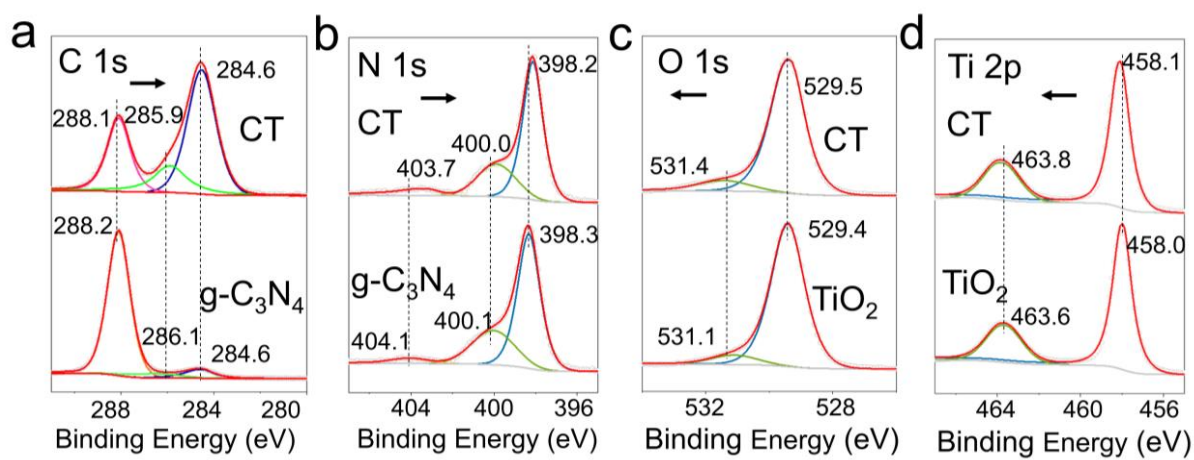

**Figure 4.** XPS spectra of the as-fabricated samples :(a) C 1s (b)N 1s (c) O 1s (d) Ti 2p.

Thermogravimetric analysis (TGA) reveals that g-C$_3$N$_4$ undergoes significant weight loss at ~600 °C, corresponding to its thermal decomposition (Figure S2, Supporting Information), whereas TiO$_2$(B) exhibits excellent thermal stability under the same conditions. For CT, a weight loss of approximately 40% is observed, which is consistent with the decomposition of the g-C$_3$N$_4$ component and aligns with the initial precursor ratio used for synthesis (100 mg TiO$_2$(B) and 70 mg g-C$_3$N$_4$, corresponding to 41 wt.% g-C$_3$N$_4$) [33]. Nitrogen adsorption–desorption measurements indicate that TiO$_2$(B) has a specific surface area of 24.4 m²·g⁻¹ (Figure

S3, Supporting Information). After coupling with g-C₃N₄, the surface area of CT increases to 39.1 m²·g⁻¹, which is higher than that of TiO₂(B) but lower than that of pristine g-C₃N₄ [29]. The increased surface area in CT provides additional catalytic sites, potentially enhancing photocatalytic performance.

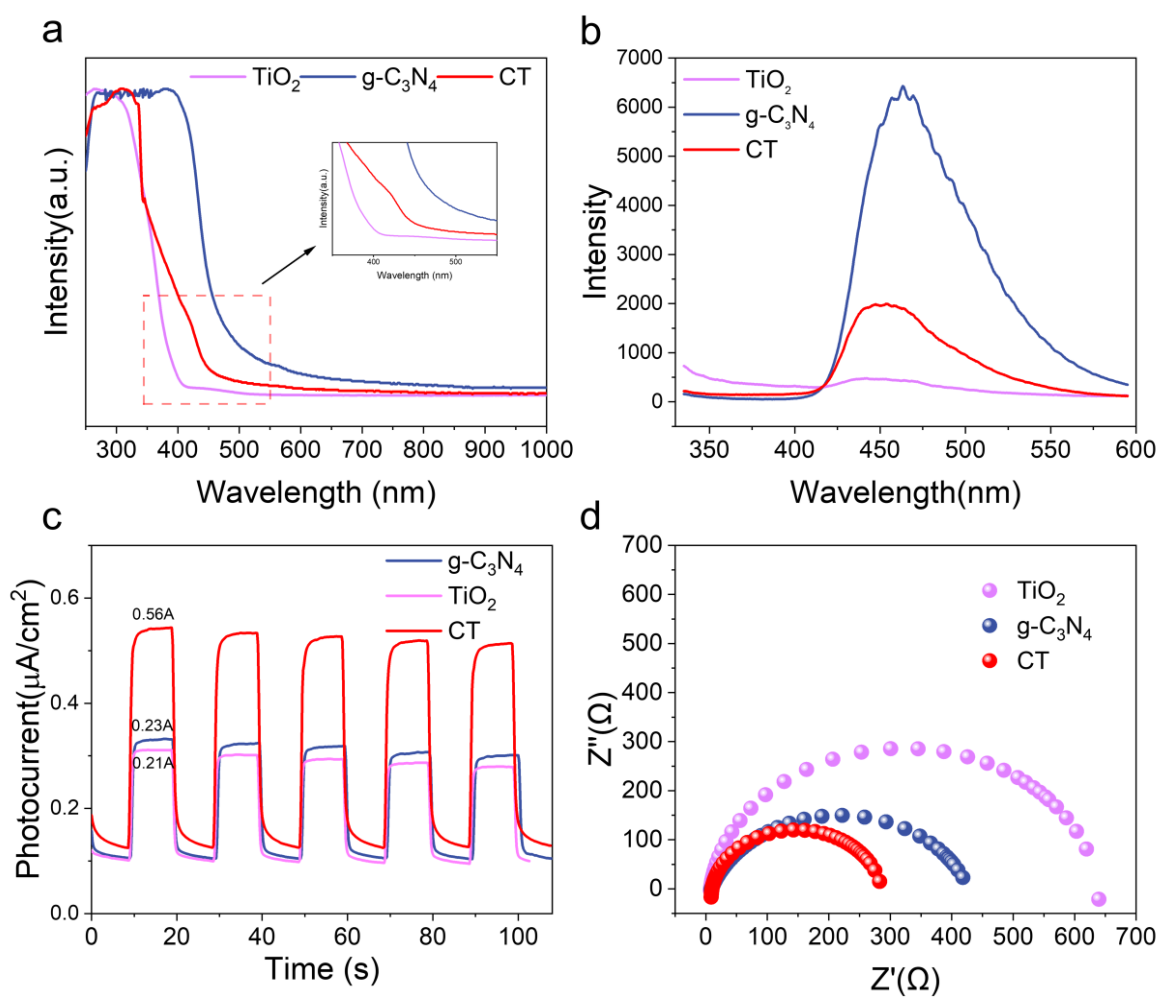

**Figure 5.** (a) UV–vis absorption spectra and (b) steady-state PL spectra under 315 nm excitation, (c) transient photocurrent responses of CT, TiO₂, and g-C₃N₄, (d) electrochemical impedance spectra of CT, TiO₂, and g-C₃N₄.

The UV–vis absorption spectra of TiO₂(B), g-C₃N₄, and CT are shown in Figure 5a. The absorption edge of TiO₂(B) is located at ~410 nm, consistent with its wide bandgap (3.0–3.2 eV) [34], while g-C₃N₄ exhibits a red-shifted edge at ~470 nm due to its intrinsic visible-light response (bandgap 2.6–2.7 eV) [35]. The CT heterojunction shows a hybrid absorption profile with an edge positioned between those of the individual components, indicating that heterojunction formation extends the light absorption range and enhances visible-light harvesting [35]. Photoluminescence (PL) spectra were employed to probe the recombination of photogenerated charge carriers. Pristine g-$C_3N_4$ exhibits a strong emission peak at ~460 nm under 315 nm excitation, whereas $TiO_2$(B) shows only a weak emission at ~450 nm due to its indirect bandgap [36]. In CT, the PL intensity at ~460 nm is quenched by ~70%, which can be attributed to interfacial charge transfer from g-$C_3N_4$ to $TiO_2$(B), consistent with the 60 wt.% $TiO_2$(B) composition determined by TGA. This quenching indicates effective suppression of electron–hole recombination in g-$C_3N_4$ [37]. Transient photocurrent measurements under simulated sunlight (five light on/off cycles) show stable responses for both $TiO_2$(B) (0.21 mA·cm$^{-2}$) and g-$C_3N_4$ (0.23 mA·cm$^{-2}$). Notably, the photocurrent density of CT is approximately twice that of the single-component materials, confirming enhanced separation of photogenerated electron–hole pairs and improved photo-to-electron conversion efficiency [38]. Electrochemical impedance spectroscopy (EIS) further corroborates this conclusion. The Nyquist plot of CT exhibits a smaller semicircle than $TiO_2$(B) and g-$C_3N_4$, indicating reduced charge-transfer resistance and more efficient interfacial electron transport in the heterojunction [39]. Collectively, these results demonstrate that the g-$C_3N_4$/$TiO_2$(B) S-scheme heterojunction effectively promotes light absorption, facilitates charge separation, and enhances photocatalytic performance.

Mott-Schottky (M-S) measurements were performed to determine the flat-band potentials of the photocatalysts. The results indicate that g-$C_3N_4$ and $TiO_2$(B) exhibit flat-band potentials

of –1.39 and –0.59 V versus Ag/AgCl, respectively (Figure S4). Using the equation ($E_{fb(NHE)} = E_{fb(pH=0,\ Ag/AgCl)} + E_{Ag/AgCl} + 0.059 \times pH$), with $E_{Ag/AgCl} = 0.197$ V and pH = 6.8, the flat-band potentials relative to the normal hydrogen electrode (NHE) were calculated to be -0.8 and 0 V for g-$C_3N_4$ and $TiO_2$(B), respectively [40]. Considering that the flat-band potential is 0.3 V more positive than the CB potential, the CB potentials of g-$C_3N_4$ and $TiO_2$(B) were derived as -1.1 and -0.3 V (vs. NHE), respectively [27]. Based on the UV-vis absorption spectra, the bandgaps of g-$C_3N_4$ and $TiO_2$(B) are 2.63 and 3.02 eV, corresponding to VB potentials of 1.53 and 2.72 V, respectively. Theoretical calculations reveal that the work function ($W_F$) of g-$C_3N_4$ (~3.98 eV) is smaller than that of $TiO_2$(110) (4.75 eV) (Figure 6a-b). Differential charge density mapping at the g-$C_3N_4$/$TiO_2$(B) hetero-interface shows significant charge redistribution upon contact, with electron accumulation on TiO2(B) and hole accumulation on g-C3N4. Bader charge analysis confirms that this interfacial electron transfer originates from the difference in $E_f$ and interfacial bonding interactions between g-$C_3N_4$ and $TiO_2$(B) [41].

Integrating the M–S measurements, UV–vis data, and theoretical calculations, the energy band alignment of g-$C_3N_4$ and $TiO_2$(B) is illustrated in Figure 6d. The VB of $TiO_2$(B) and the CB of g-$C_3N_4$ are suitable for oxidizing water to •OH radicals and reducing water to $H_2$, respectively [42]. Due to the lower WF of g-$C_3N_4$, interfacial electron transfer occurs from g-$C_3N_4$ to $TiO_2$(B), generating an internal electric field directed from g-$C_3N_4$ to $TiO_2$(B). This induces upward and downward band bending at the energy edges of g-$C_3N_4$ and $TiO_2$(B), respectively. Upon light excitation, photogenerated electrons in the CB of $TiO_2$(B) migrate across the heterointerface to the VB of g-$C_3N_4$, where they recombine with holes. Consequently, highly reducing electrons remain in the CB of g-$C_3N_4$, and strongly oxidizing holes are retained in the VB of $TiO_2$(B) [43]. This S-scheme charge-transfer mechanism enhances the redox capability of the heterojunction and prolongs the lifetime of photogenerated charge carriers, consistent with the PL, photocurrent, and EIS results.

To further validate the proposed S-scheme mechanism, electron paramagnetic resonance (EPR) spectroscopy was performed on the as-synthesized samples. As shown in Figure 6e, signals corresponding to superoxide radicals (•$O_2^-$) are observed for all samples. The intensity of the •$O_2^-$ signal in g-$C_3N_4$ is much higher than that of $TiO_2$(B), which can be attributed to the more negative CB potential of g-$C_3N_4$ (–1.1 V vs. NHE) relative to the $O_2$/•$O_2^-$ redox potential (–0.33 V vs. NHE). In contrast, the CB potential of $TiO_2$(B) is comparable to the $O_2$/•$O_2^-$ potential, resulting in weak •$O_2^-$ signal intensity. Notably, the •$O_2^-$ signal in CT exceeds that of pristine g-$C_3N_4$, indicating enhanced separation efficiency of photogenerated electron–hole pairs in the heterojunction. Figure 6f shows the EPR spectra for hydroxyl radicals (•OH). A distinct •OH signal is observed in $TiO_2$(B), whereas g-$C_3N_4$ exhibits no detectable •OH, consistent with its VB potential being lower than the $OH^-$/•OH redox potential (2.69 V vs. NHE), which prevents oxidation of $OH^-$. Importantly, the •OH signal in CT is stronger than that of pure $TiO_2$(B), despite the presence of g-$C_3N_4$, which cannot generate •OH independently. These EPR results provide direct evidence for the formation of the g-$C_3N_4$/$TiO_2$(B) S-scheme heterojunction. The heterojunction promotes interfacial charge transfer, efficiently separating photogenerated electrons and holes while maintaining strong redox potentials [44].

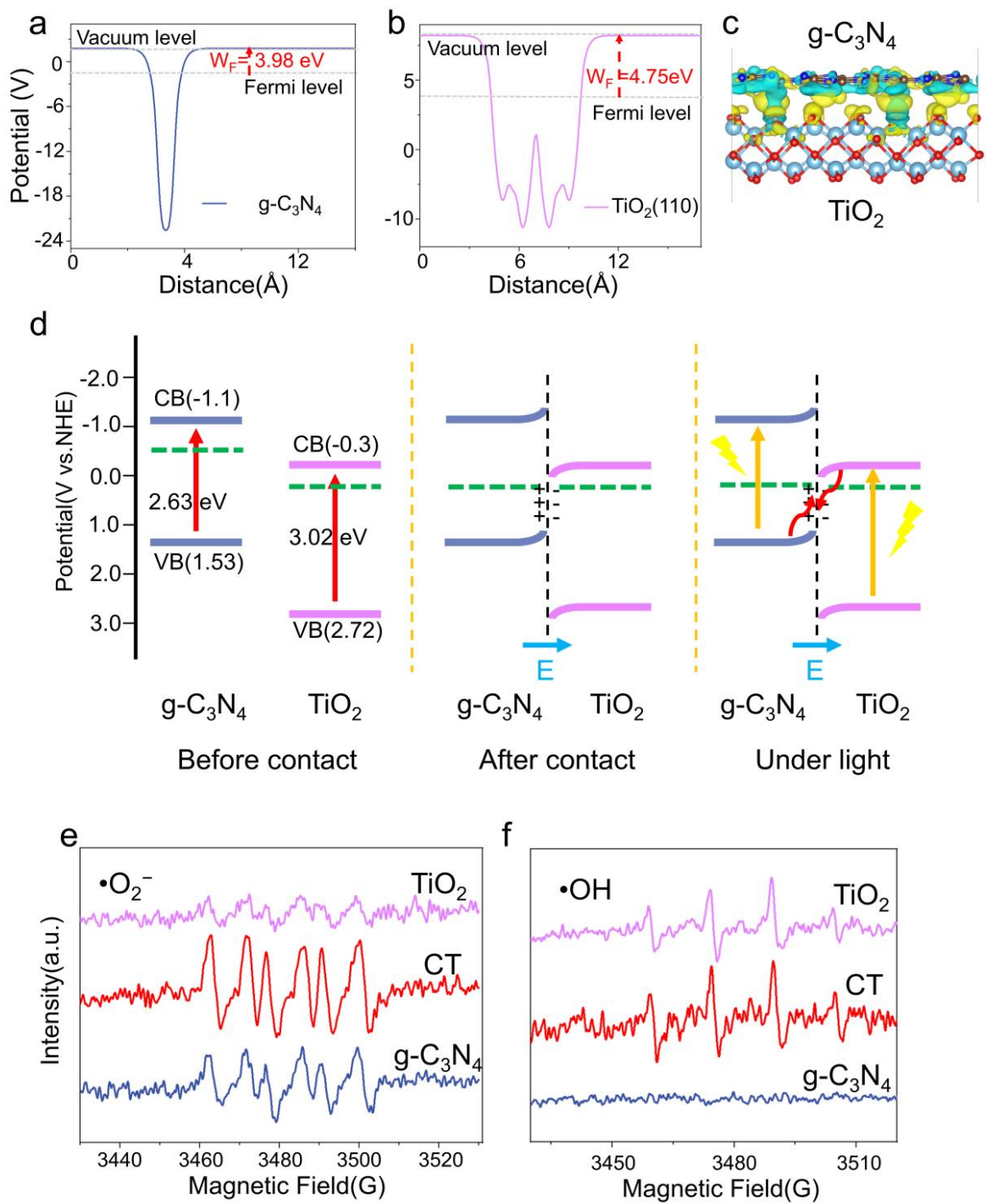

**Figure 6.** (a) Work function of $TiO_2$, (b) Work function of $g-C_3N_4$, (c) Charge density different of $g-C_3N_4/TiO_2$ interface, (d) Photocatalytic mechanism of $g-C_3N_4/TiO_2$ heterojuntion, (e) EPR Spectra of ·$O_2^-$ Radicals and (f) EPR Spectra of ·OH Radicals.

The photocatalytic $H_2$ evolution activities of the as-synthesized samples under simulated sunlight were first evaluated using water photoreduction with methanol as a sacrificial agent and Pt as a cocatalyst. As shown in Figure 7a, the $H_2$ evolution rates of pristine g-$C_3N_4$ and $TiO_2$(B) are ~0.97 and ~1.3 mmol·g$^{-1}$·h$^{-1}$, respectively. Upon coupling g-$C_3N_4$ with $TiO_2$(B), the $H_2$ evolution rate of CT increases to 1.98 mmol·g$^{-1}$·h$^{-1}$, which is approximately 1.5 and 2.0 times higher than those of the individual components. This enhancement confirms that the S-scheme charge-transfer mechanism effectively promotes the separation of photogenerated electron–hole pairs, thereby boosting photocatalytic performance.

To optimize the heterojunction composition, a series of control samples with varying g-$C_3N_4$ contents were synthesized. XRD, FT-IR, PL, and UV–vis characterizations confirm the successful formation of heterojunctions with different g-$C_3N_4$ loadings (Figure S5, Supporting Information). The samples prepared with 50, 60, 80, and 90 mg of g-$C_3N_4$ precursor are denoted as CT-50, CT-60, CT-80, and CT-90, respectively. As shown in Figure 7b, the $H_2$ evolution rate depends on the g-$C_3N_4$ content, reaching a maximum at ~40 wt.% (70 mg precursor). When the g-$C_3N_4$ content exceeds this value, the $H_2$ production decreases, likely due to excessive g-$C_3N_4$ covering $TiO_2$(B) surfaces and reducing the accessibility of active interfacial sites, which hinders charge separation and transfer. The long-term stability of the optimal CT photocatalyst was evaluated by cycling tests under continuous simulated sunlight. As shown in Figure 7c, the $H_2$ evolution rate remains nearly constant over 18 h, with only a slight decline during the final two hours of a 20 h test. Moreover, CT maintains stable performance over 10 consecutive cycles, demonstrating its excellent photostability and durability.

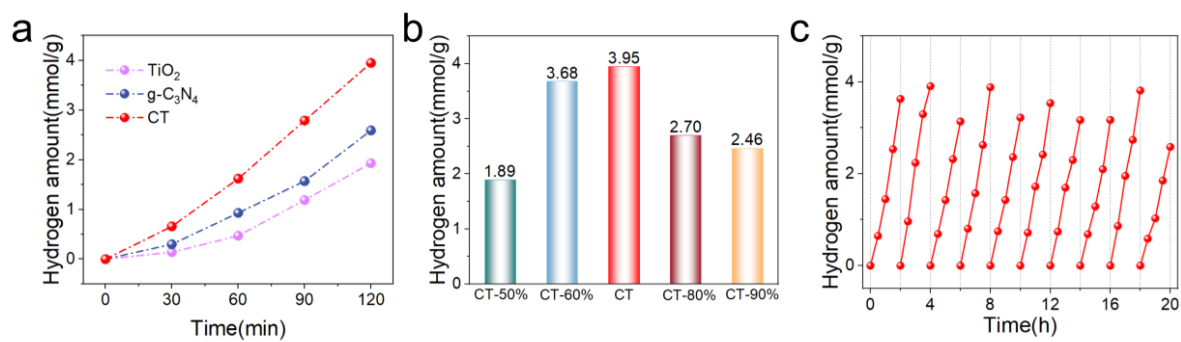

**Figure 7** (a) Photocatalytic hydrogen evolution performance of TiO$_2$, g-C$_3$N$_4$ and CT, (b) photocatalytic hydrogen evolution rates of CT, (c) recyclability and stability of CT in photocatalytic hydrogen evolution.

To evaluate the versatility of the photocatalyst, systematic experiments were conducted to examine its ability to simultaneously degrade organic pollutants and evolve H$_2$ under simulated sunlight. Given its superior H$_2$ evolution performance, CT was selected as the model photocatalyst. Six representative pollutants—including amoxicillin (AMX), ciprofloxacin (CIP), ofloxacin (OFLX), tetracycline (TC), rhodamine B (RhB), and methyl orange (MO) (Figure 8a)—were employed to assess photocatalytic degradation (Figure 8b). After 90 min of simulated sunlight irradiation, CT achieved removal efficiencies of 98.2%, 89.4%, 96.1%, 85.7%, 83.8%, and 97.8% for AMX, CIP, OFLX, TC, RhB, and MO, respectively. Simultaneously, corresponding H$_2$ evolution yields of ~20.70, ~6.53, ~3.69, ~3.51, ~1.38, and ~1.14 μmol·g$^{-1}$ were observed (Figure 8c). The differences in coupled degradation–H$_2$ evolution performance are attributed to variations in the molecular size and structure of the pollutants. The performance of CT was also benchmarked against other reported heterostructured photocatalysts (Table S1, Supporting Information). The cyclic stability of CT was evaluated over eight consecutive cycles (16 h total) for AMX degradation (Figure 8d). After eight cycles, CT retained over 86% degradation efficiency, demonstrating excellent durability. XPS characterization of the recovered photocatalyst (Figure S6, Supporting

Information) showed no significant changes in the chemical states of Ti, O, C, or N, confirming that CT maintains structural and chemical stability during photocatalysis. These results demonstrate that the g-$C_3N_4$/$TiO_2$(B) S-scheme heterojunction enables efficient interfacial charge separation, providing strong oxidizing holes and reducing electrons that simultaneously drive organic pollutant degradation and $H_2$ evolution.

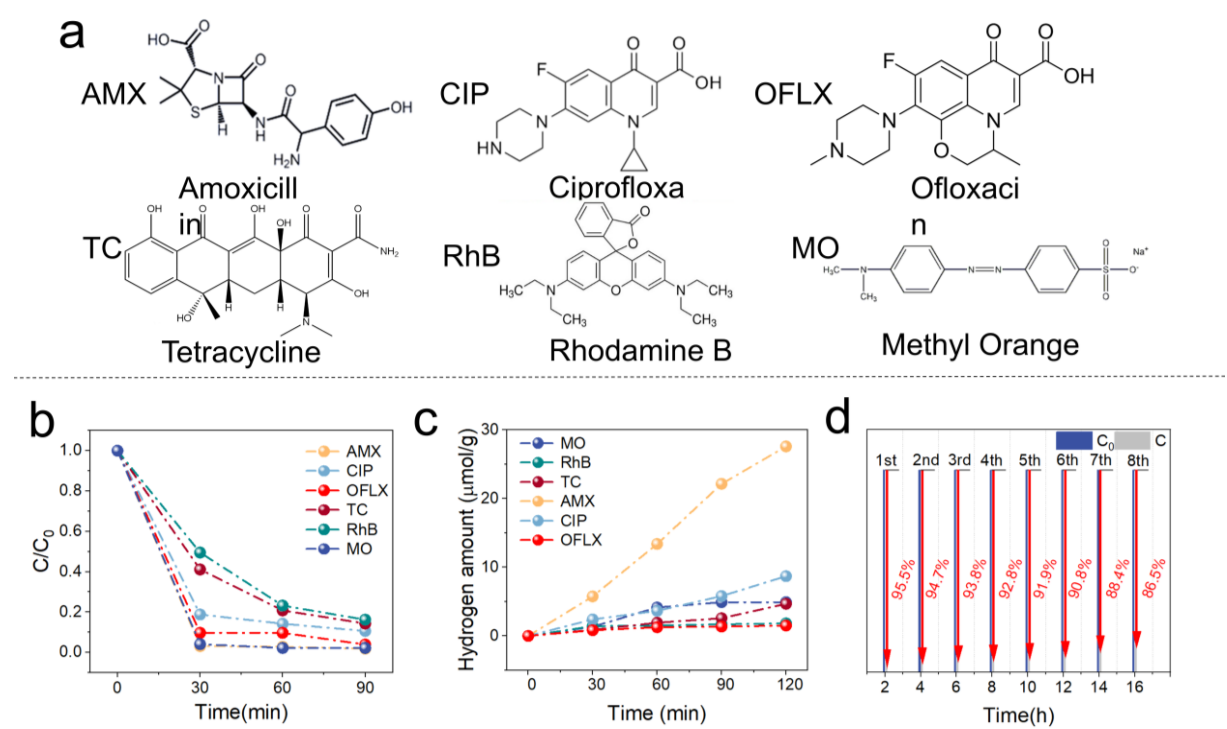

**Figure 8.** (a) Structural formulas of six representative pollutants, (b) photocatalytic degradation of pollutants, (c) photocatalytic hydrogen evolution, (d) cycling stability in photocatalytic degradation.

To elucidate the reaction intermediates during AMX degradation over CT under simulated sunlight, in-situ FT-IR spectroscopy was employed. As shown in Figure S7, five distinct peaks emerge, with intensities increasing over irradiation time. A peak at ~1455 cm$^{-1}$ corresponds to bicarbonate ($HCO_3^-$) intermediates, while the peak at ~1582 cm$^{-1}$ is attributed to carboxyl-containing species ($COOH^-$). The peak near ~1645 cm$^{-1}$ is assigned to C=O stretching, indicating oxidation of AMX to form carbonyl-containing intermediates. Notably, a peak at

~2368 cm$^{-1}$, corresponding to CO$_2$, gradually intensifies with reaction time, confirming the progressive mineralization of AMX into CO$_2$ [45].

Liquid chromatography-mass spectrometry (LC-MS) was employed to systematically elucidate the degradation mechanism of AMX. By analyzing the degradation by-products at different irradiation time points (0, 30, 60, and 120 min), the detailed degradation pathways of AMX were proposed in Figure 9a. We demonstrated the following three pathways:

**Pathway I:** The β-lactam ring of AMX opens to form intermediate A1, which undergoes decarboxylation to yield A2. A2 then undergoes C–C bond cleavage and deamination to generate A3. Subsequent molecular rearrangement and cleavage of C–H and C–N bonds transform A3 into A4. Finally, ring opening of A4 via C–N and C–C bond cleavage, accompanied by carboxyl group dissociation, produces A5 [46].

**Pathway II:** Cleavage of AMX's central peptide bond produces A6 and A7, which degrade independently. For A6, C-N bond cleavage on its internal lactam ring forms A8; subsequent five-membered ring opening with C-N and C-S bond cleavage generates A10. For A7, amino group dissociation first forms A9, followed by carboxyl group dissociation to yield A11 [47].

**Pathway III:** Similar to Pathway I, this pathway initiates with AMX lactam ring opening to form A1. A1 then loses a hydroxyl group to generate A2, which undergoes C-N and C-S bond cleavage of its five-membered ring to form A12. Subsequent cleavage of A12's central peptide bond produces A13, which is finally oxidized with carboxyl group cleavage to form A14 [48].

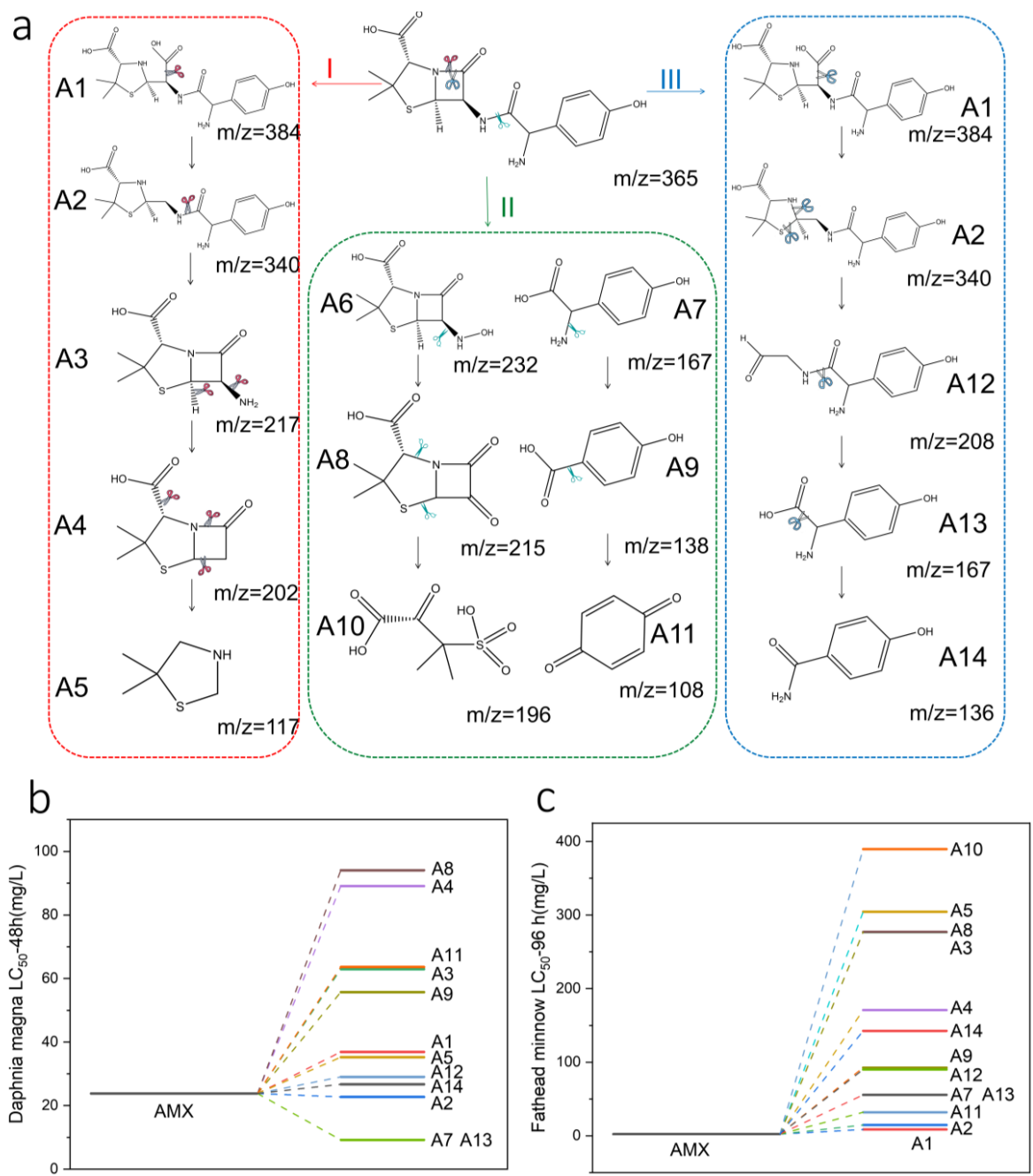

**Figure 9.** (a) Possible degradation pathways of AMX based on LC-MS analysis, (b) Acute toxicity $LC_{50}$-48h of Daphnia magna, (c) $LC_{50}$-96h of fathead minnow.

The LC–MS results indicate that active moieties of AMX, including the β-lactam ring, side-chain amino group, and benzene ring, are preferentially attacked by •OH radicals and photogenerated holes in CT. The intermediates are gradually mineralized to $CO_2$. Furthermore, toxicity simulation based on TEST/QSAR models shows elevated $LC_{50}$ values for Daphnia magna and fathead minnow after photocatalytic treatment, confirming the reduced toxicity of the treated effluent [49]. For other pollutants (CIP, OFLX, TC, RhB, and MO), their conjugated double bonds and polycyclic structures render them susceptible to radical-induced bond cleavage or ring opening, ultimately leading to mineralization into $CO_2$. These findings confirm that the S-scheme g-$C_3N_4$/$TiO_2$(B) heterojunction not only enhances charge separation but also provides highly oxidative holes and reducing electrons that drive efficient degradation and mineralization of complex organic pollutants.

## 4. Conclusions

In summary, g-$C_3N_4$/$TiO_2$(B) S-scheme heterojunctions were successfully synthesized, providing an effective platform for the simultaneous photocatalytic degradation of organic pollutants and $H_2$ production under simulated sunlight. Among the samples, the optimal heterojunction containing 40 wt.% g-$C_3N_4$ exhibited superior photocatalytic performance compared to pristine g-$C_3N_4$ and $TiO_2$(B). Specifically, it achieved an $H_2$ evolution rate of 1.98 mmol·g$^{-1}$·h$^{-1}$, which is 1.5- and 2.0-fold higher than those of pure g-$C_3N_4$ and $TiO_2$(B), respectively. In parallel, the heterojunction demonstrated excellent broad-spectrum photocatalytic activity toward various organic pollutants, achieving removal efficiencies of 98.2%, 89.4%, 96.1%, 85.7%, 83.8%, and 97.8% for amoxicillin, ciprofloxacin, ofloxacin, tetracycline, rhodamine B, and methyl orange, respectively, within 90 min of simulated sunlight irradiation. Notably, $H_2$ production was simultaneously realized during pollutant degradation, with AMX degradation corresponding to the highest $H_2$ yield of ~20.70 μmol·g$^{-1}$. These results confirm that the g-$C_3N_4$/$TiO_2$(B) S-scheme heterojunction efficiently promotes interfacial charge separation, retaining strongly reducing electrons and oxidizing holes that drive both $H_2$ evolution and pollutant mineralization. This study provides a rational strategy for designing dual-functional S-scheme heterojunction photocatalysts, offering promising solutions for integrated environmental remediation and renewable energy generation.

## Acknowledgements

This research was supported by the Liaoning Revitalization Talents Program, China (XLYC1807170), and the Liaoning BaiQianWan Talents Program.

# Supplementary Material

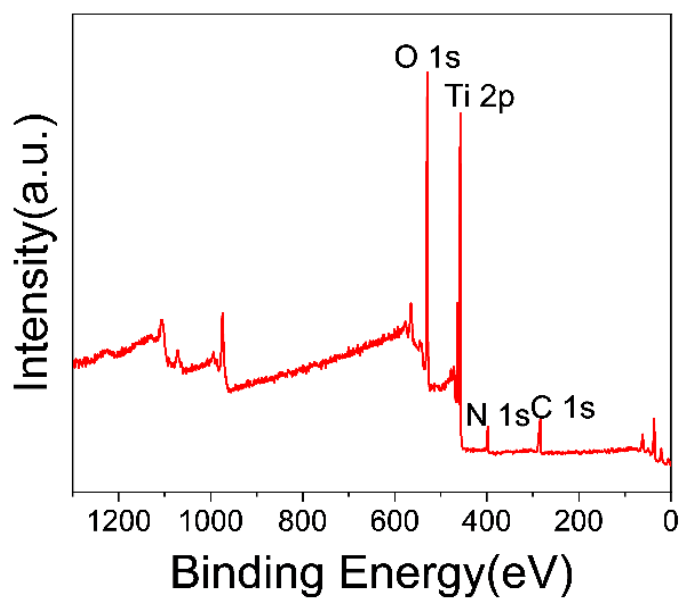

**Figure. S1**. The full XPS survey spectrum confirms the co-existence of Ti, O, C, and N element signals, verifying the successful incorporation of these four elements into the composite

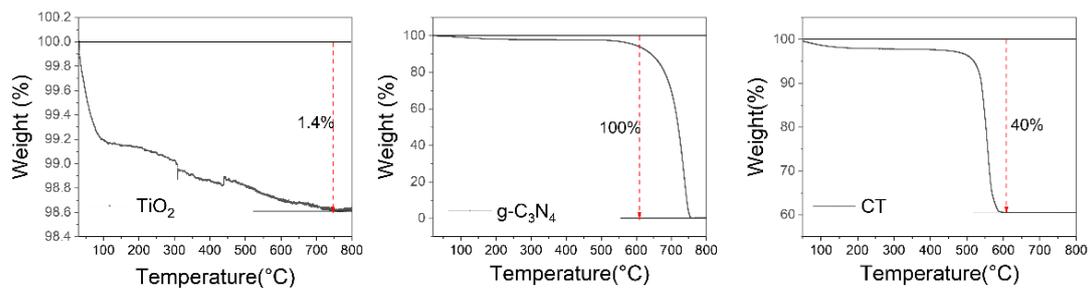

**Figure. S2**. Thermogravimetric analysis (TGA) results reveal s significant weight loss of g-$C_3N_4$ at 600℃, which is attributed to its thermal decomposition. In contrast, $TiO_2$(B) exhibits good thermal stability during the high-temperature measurement. For CT, a ~40% weight loss is detected, corresponding to the thermal decomposition of the g-$C_3N_4$ component in the heterojunction

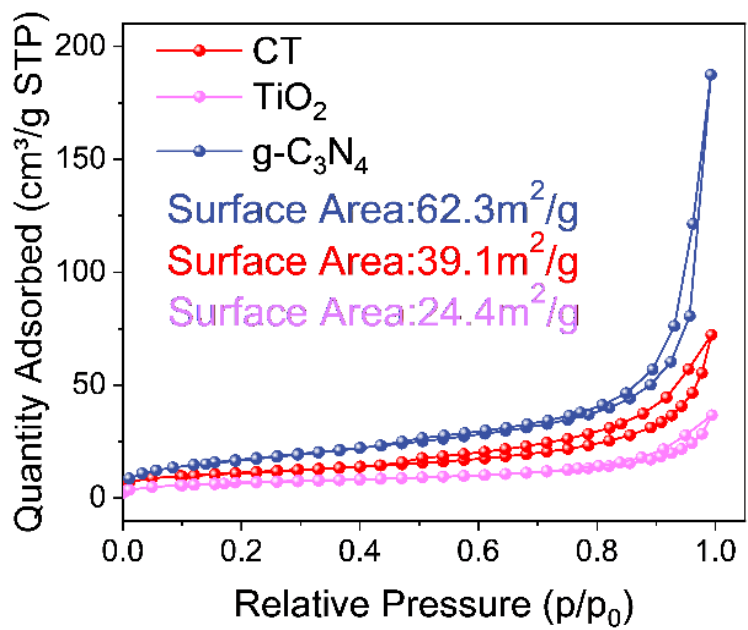

**Figure S3** Nitrogen adsorption-desorption isotherm measurements show that the specific surface area of TiO$_2$(B) is 24.4 m$^2$ g$^{-1}$. After coupling TiO$_2$(B) with g-C$_3$N$_4$, the specific surface area of the resulting heterojunction reaches 39.1 m$^2$ g$^{-1}$. In theory, larger specific surface area can provide more catalytic active sites, thereby enhancing the photocatalytic activity of the material.

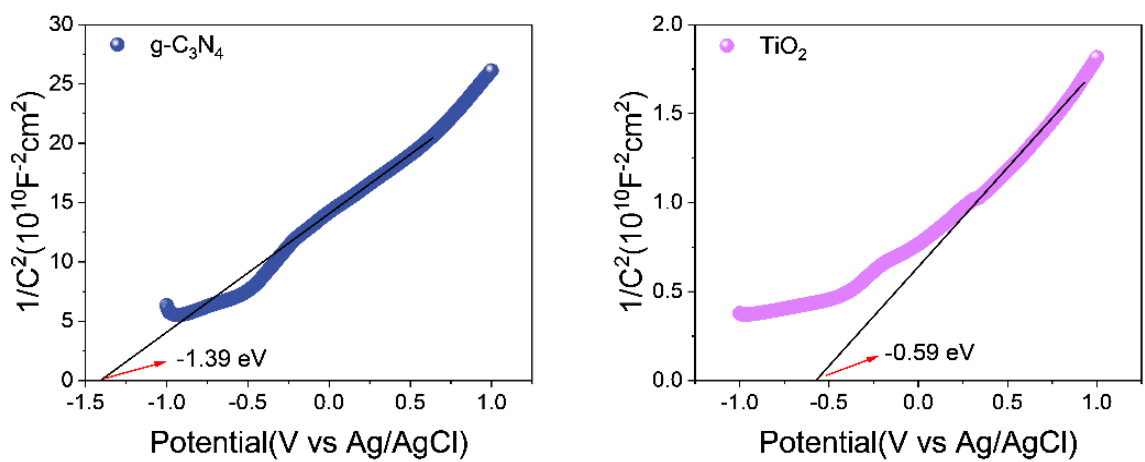

**Figure S4** Mott-Schottky (M-S) measurements were performed to determine the flat-band potentials of the photocatalysts, and the results indicate that the g-$C_3N_4$ and $TiO_2$(B) are -1.39 and -0.59 V (vs. Ag/AgCl)

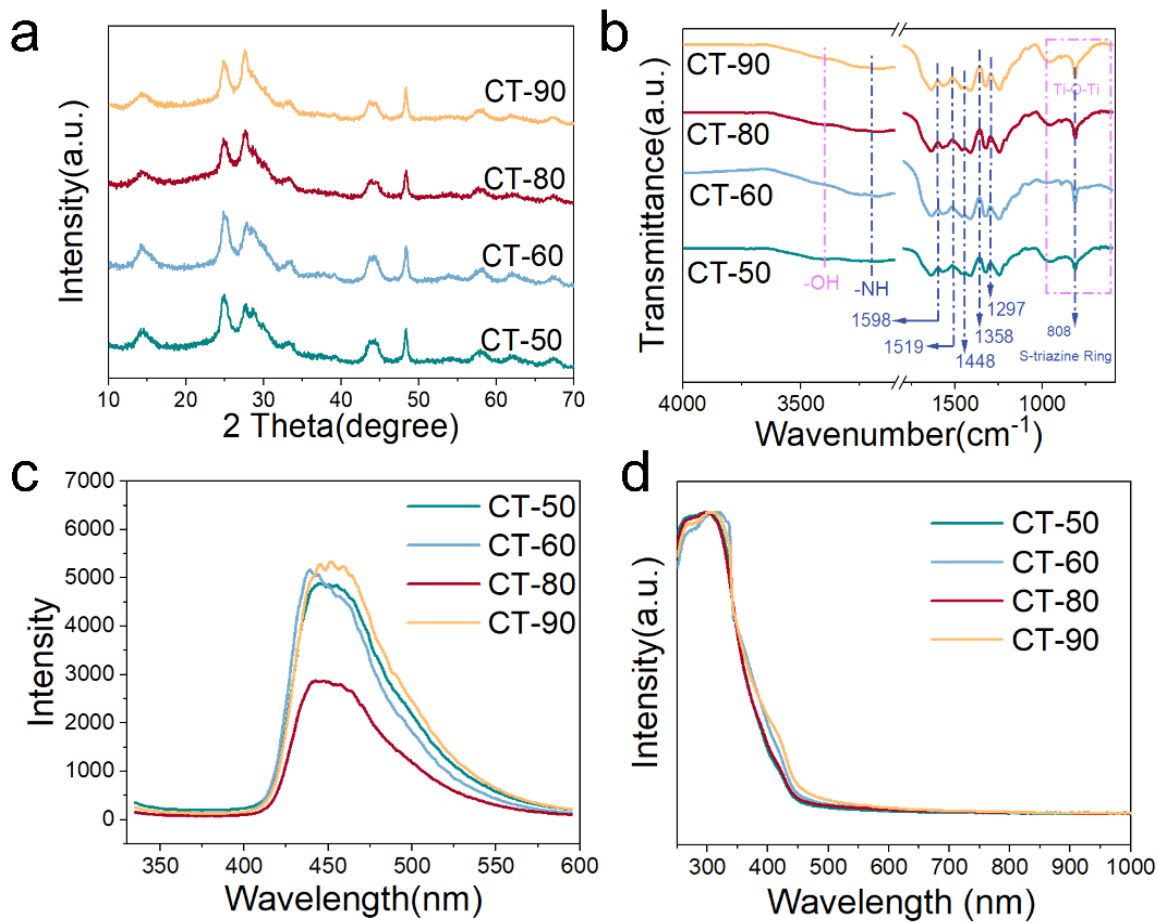

**Figure S5.** To further validate this result, a series of control samples were synthesized by adjusting the dosage of g-C$_3$N$_4$ in the g-C$_3$N$_4$/TiO$_2$(B) heterojunction. XRD, FT-IR, PL, UV-visible absorption spectra characterizations confirm the successful synthesis of heterojunctions with different g-C$_3$N$_4$ contents. The control samples synthesized with 50, 60, 80, and 90 mg of g-C$_3$N$_4$ precursor are denoted as CT-50, CT-60, CT-80, and CT-90, respectively

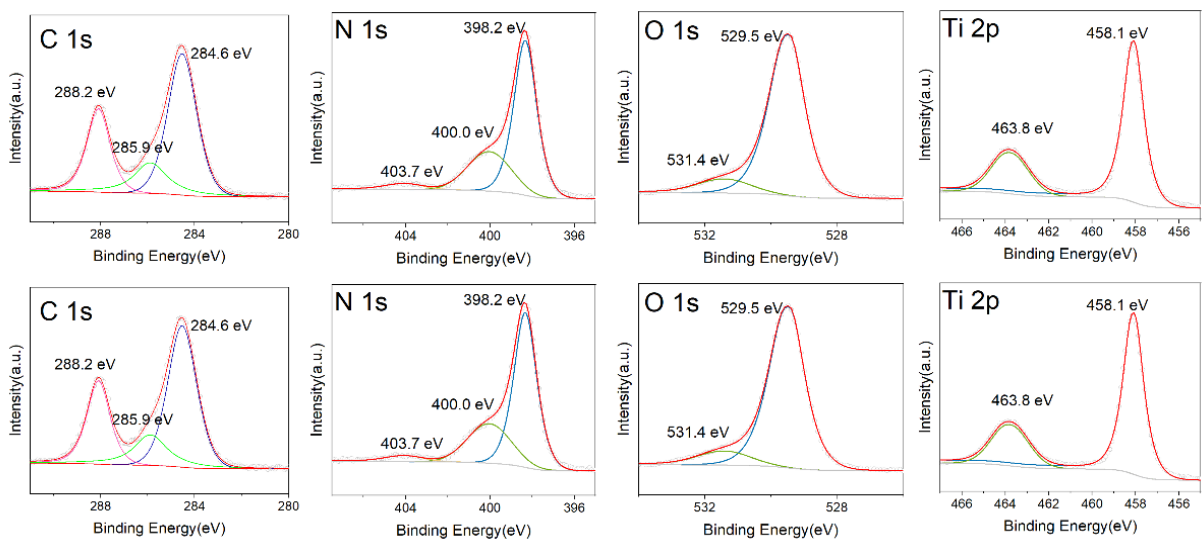

**Figure S6.** The recovered photocatalyst was re-characterized by XPS. No obvious changes in the chemical environments of the atoms were observed before and after the reaction, confirming that the composite retains structural and chemical stability during photocatalysis

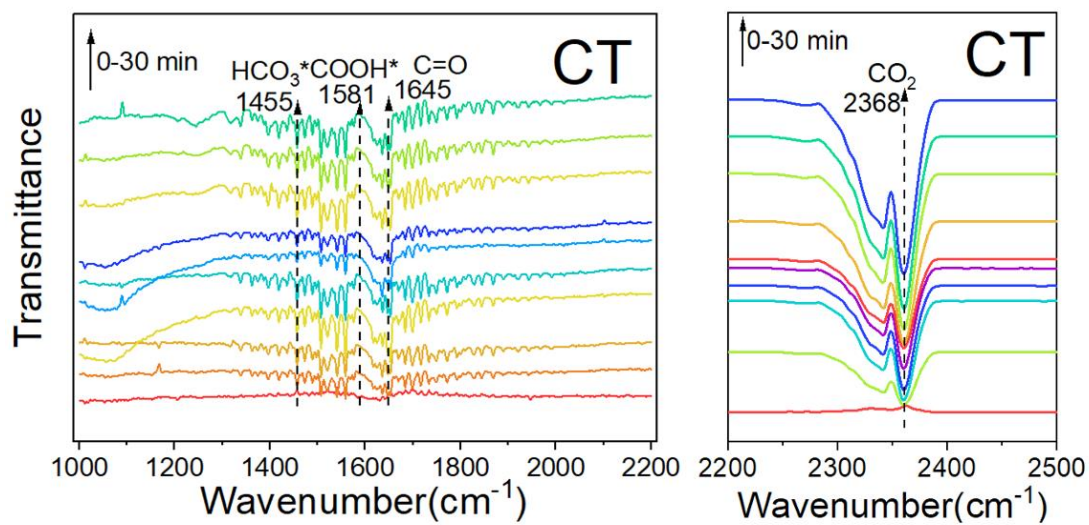

**Figure S7.** In-situ Fourier transform infrared spectroscopy (in-situ FT-IR) was used for characterization. Five distinct characteristic peaks appeared in the in-situ IR spectra, and their intensities gradually increased with prolonged light irradiation time. These results confirm that carbon dioxide ($CO_2$) is a product of the photodegradation of amoxicillin (AMX), indicating that the target pollutant has been efficiently mineralized.

**Table S1** This article compares hydrogen production and degradation with other relevant literature.

The comparison of hydrogen production volume and degradation efficiency of this paper with those in other literature

| Ref | $H_2$ rate($\mu mol \cdot g^{-1} \cdot h^{-1}$) | contaminant | degradation time | pollutant removal rate | Photocatalysts |
|---|---|---|---|---|---|
| [1] | 1620 | -- | -- | -- | d-$Ti_3C_2$/$TiO_2$/g-$C_3N_4$ |
| [2] | 317 | -- | -- | -- | $N_2$S-$TiO_2$/g-$C_3N_4$ |
| [3] | 181 | -- | -- | -- | MoS@$TiO_2$ |
| [4] | 1150 | -- | -- | -- | C-$TiO_2$/g-$C_3N_4$ |
| [5] | 985.8 | -- | -- | -- | $WS_2$/C-$TiO_2$/g-$C_3N_4$ |
| [6] | -- | Methylene Blue (MB) | 120 min | 65% removal rate | S-$TiO_2$/g-$C_3N_4$ |
| [7] | -- | Tetracycline (TC) | 120 min | 94.8% removal rate | N-$TiO_2$ |
| [8] | 102.4 | Tetracycline (TC) | 120 min | 52.16% removal rate | CIS/CN |
| [9] | 1402.7 | -- | -- | -- | g-$C_3N_4$/$BiO_{1.2}I_{0.6}$ |
| [10] | 555.8 | Methyl Orange (MO) | 180 min | 95% removal rate | b-$TiO_2$/g-$C_3N_4$ |
| [11] | 910 | Methyl Orange (MO) | 80 min | 99.3% removal rate | Ag/$TiO_2$ |
| [12] | 8931.3 | -- | -- | -- | GCN/NT NFs |
| [13] | 1975 | Amoxicillin (AMX) | 90 min | 98.2% removal rate | this work |